\documentclass[aps,prb,twocolumn,nofootinbib,superscriptaddress]{revtex4-2}

\usepackage[utf8]{inputenc}
\usepackage[T1]{fontenc}

\usepackage{amsmath}
\usepackage{amsfonts}
\usepackage{amssymb}
\usepackage{graphicx}
\usepackage{dsfont}
\usepackage{mathtools}

\usepackage[english]{babel}

\usepackage{physics}

\usepackage{hyphenat}
\usepackage{hyperref}
\usepackage{xcolor}
\usepackage{xspace}
\usepackage{placeins}

\newcommand{\ie}{i.e.\,}
\newcommand{\eg}{e.g.\,}

\usepackage[
    acronym,
    hyperfirst = false
]{glossaries}
\glsdisablehyper 
\newacronym{cst}{CST}{continuous similarity transformation}
\newacronym{cut}{CUT}{continuous unitary transformation}
\newacronym{mft}{MFT}{mean-field theory}
\newacronym{lswt}{LSWT}{linear spin-wave theory}
\newacronym{qpc}{qpc}{quasi-particle conserving}
\newacronym{ins}{INS}{inelastic neutron scattering}
\newacronym{pbc}{pbc}{periodic boundary conditions}
\newacronym{apbc}{apbc}{antiperiodic boundary conditions}
\newacronym{rod}{ROD}{residual off-diagonality}

\newcommand{\dsc}{d_\text{sc}}

\usepackage[capitalise]{cleveref}
\crefname{section}{Sec.}{sections}
\crefname{appendix}{App.}{appendizes}

\begin{document}

\title{Quantum decay of magnons in the unfrustrated honeycomb Heisenberg model}
\author{Calvin Krämer}
\affiliation{Department of Physics, Staudtstra{\ss}e 7, Friedrich-Alexander-Universit\"at Erlangen-N\"urnberg (FAU), D-91058 Erlangen, Germany}
\author{Dag-Björn Hering}
\affiliation{Condensed Matter Theory, Department of Physics, TU Dortmund University, Otto-Hahn-Stra\ss{}e 4, 44227 Dortmund, Germany}
\author{Vanessa Sulaiman}
\affiliation{Condensed Matter Theory, Department of Physics, TU Dortmund University, Otto-Hahn-Stra\ss{}e 4, 44227 Dortmund, Germany}
\author{Matthias R. Walther}
\affiliation{Department of Physics, Staudtstra{\ss}e 7, Friedrich-Alexander-Universit\"at Erlangen-N\"urnberg (FAU), D-91058 Erlangen, Germany}
\author{Götz S. Uhrig}
\affiliation{Condensed Matter Theory, Department of Physics, TU Dortmund University, Otto-Hahn-Stra\ss{}e 4, 44227 Dortmund, Germany}
\author{Kai Phillip Schmidt}
\affiliation{Department of Physics, Staudtstra{\ss}e 7, Friedrich-Alexander-Universit\"at Erlangen-N\"urnberg (FAU), D-91058 Erlangen, Germany}

\begin{abstract}
We investigate the physical properties of elementary magnon excitations of the ordered antiferromagnetic Heisenberg model on the honeycomb lattice using quantum Monte Carlo (QMC) simulations, series expansions (SE), and continuous similarity transformations (CST). 
The stochastic analytic continuation method is used to determine the dynamic structure factor from correlation functions in imaginary time obtained by QMC. 
In contrast to the ``roton minimum'' of the square lattice Heisenberg antiferromagnet, we find that magnons on the honeycomb lattice completely decay in the corner of the Brillouin zone ($K$-point);  the entire weight is shifted into the continuum.
These findings are fully supported by SE and CST in momentum space.
The extrapolated one-magnon dispersion obtained from SE about the Ising limit quantitatively agrees with the extracted QMC excitation energies except around the $K$-point, where large uncertainties in the extrapolation indicate the magnon decay.
This quantum decay is further confirmed and understood by the CST, which yields a divergent flow when enforcing a magnon quasi-particle picture.
The divergence originates from strong attractive magnon-magnon interactions
leading to a bound state and thereby to a three-magnon continuum overlapping with 
the one-magnon state.
This has the magnon quasi-particle picture break down at high energies on the honeycomb lattice.
\end{abstract}

\maketitle

\section{Introduction}
The antiferromagnetic Heisenberg model is a central model in quantum magnetism, describing a wide range of materials, including the undoped high $T_c$ cuprate superconductors \cite{Anderson1987, Manousakis1991}.
On bipartite lattices with dimension $d \geq 2$ and short-range interactions, a Néel phase is found in the ground state at zero temperature. 
In one dimension, this spontaneous symmetry breaking is prohibited by a variant of the 
Mermin Wagner theorem \cite{Mermin1966, Hohenberg1967, Pitaevskii1991}.
The low-energy excitations of the Néel phase are gapless Goldstone modes \cite{Nambu1960, Goldstone1961, Goldstone1962}, referred to as magnons.
On the one hand, they are well described by linear spin-wave (LSW) theory in the semiclassical limit. On the other hand, Heisenberg models with frustrated geometries or interactions 
provide a natural setting for fractionalization (quasi particles with fractional quantum numbers) and spin-liquid behavior \cite{Oh2013, Ferrari2018, Ferrari2020, Drescher2023, Hering2024}.
Here, competition between interactions suppresses magnetic order and enhances quantum fluctuations, producing deviations from LSW predictions.

But also non-frustrated bipartite lattices show unexpected spectral features.
A prominent and well studied example is the antiferromagnetic Heisenberg model on the square lattice.
Numerical studies utilizing quantum Monte Carlo (QMC), series expansions (SE), and \gls{cst} \cite{Sandvik2001, Zheng2005, Powalski2015, Shao2017, Powalski2018, Verresen2018, Walther2023, Caci2024} consistently reveal a deviation from the LSW magnon dispersion for large frequencies at momentum $k=(\pi, 0)$.
Additionally, a pronounced reduction of the magnon intensity in the dynamic structure factor is found.
Experiments on the compound $\mathrm{Cu(DCOO)}_2 \cdot \mathrm{4D}_2\mathrm{O}$, which is well described by the antiferromagnetic Heisenberg model on a square lattice, confirm this observation \cite{Ronnow2001, Christensen2007, DallaPiazza2014}.
Significant weight is shifted to the continuum, which suppresses the magnon energy, resulting in a dip in the one-particle dispersion.
This phenomena entitled ``roton minimum'' can be attributed to strong magnon–magnon interactions between the one- and three-magnon sector \cite{Powalski2018}.
Experiments supported by theoretical calculations using Gutzwiller-projected trial wavefunctions in contrast interpret the phenomena as quantum decay of the magnon to two deconfined spinons \cite{Ho2001, DallaPiazza2014}.
Adding a four-spin interaction to the model ($J$-$Q$ model) rapidly suppresses the magnon peak and leads to the interpretation that the ``roton minimum'' can also be seen as a ``precursor of deconfined criticality'' \cite{Shao2017}.

For the Heisenberg model on the honeycomb lattice, which is also bipartite and therefore unfrustrated, 
the smaller coordination number $z=3$ yields even stronger quantum fluctuations and the question arises whether the anomalies in the magnon properties seen on the square lattice become even 
more pronounced and lead ultimatively to a quantum decay of magnons.
Indeed, similar to the Heisenberg square lattice, deviations from spin wave theory results are found in the corner of the Brillouin zone ($K$-point) \cite{Maksimov2016, Ferrari2020, Gu2022, Hernandez2025}.
Experiments on the compounds $\mathrm{YbBr}_3$ and $\mathrm{YbCl}_3$ realizing the Heisenberg model on the honeycomb lattice observe broad continua and strong renormalization of magnon modes, consistent with this expectation \cite{Wessler2020, Sala2021, Sala2023, Hernandez2025}.
The key question is whether the single-magnon excitation survives across the Brillouin zone or decays entirely into multi-magnon states. 
Ref.\,\cite{Hernandez2025} hints towards a complete 
quantum decay of the magnon mode at the $K$-point.

In this work, we address this question on the honeycomb lattice using QMC, SE, and \gls{cst}.
All three of these methods were able to capture the elementary excitations 
on the closely related Heisenberg model on the square lattice very well.
We will therefore compare our methods and results for the honeycomb lattice with previous results for the square lattice throughout this paper.
The QMC simulations are complemented by stochastic analytic continuation \cite{Sandvik1998, Mishchenko2000, Vafayi2007, Beach2004, Sandvik2016, Shao2017, Shao2023} to compute the dynamical structure factor.
We restrict the sampling procedure \cite{Sandvik2016, Shao2017, Shao2023} to separate the weight of the magnon excitation from the multi-particle continuum to determine the weight on the magnon peak.
By optimizing the weight of the magnon peak for several system sizes, we observe a convergence to zero weight at the $K$-point.

The quantum decay observed by QMC is fully supported by SE and \gls{cst} in momentum space.
The extrapolated one-magnon dispersion obtained from SE about the Ising limit quantitatively 
agrees with the extracted QMC excitation energies except around the $K$-point, where large uncertainties in the extrapolation indicate the magnon decay.
Further, the flow of the \gls{cst} diverges when enforcing a magnon quasi-particle picture. 
This divergence can be traced back to strong attractive magnon-magnon interactions 
leading to binding which in turn shifts weight of three-magnon states down in energy.
As a consequence, single magnons can decay into multi-magnon states so that
the magnon quasi-particle picture breaks down at high energies in the honeycomb lattice.
This corroborates the vital importance of magnon-magnon interactions for the renormalization of magnon dynamics
in contrast to previous observations \cite{Gu2022} based on a cluster approach 
assuming that the inter-cluster interactions are of static mean-field type.

The paper is organized as follows: We introduce the model together with the three methods used in \cref{sec:model_methods}.
In \cref{sec:results} we present and discuss our results before concluding in \cref{sec:conclusion}. 

\section{Model and Methods}
\label{sec:model_methods}
In this section we introduce the Heisenberg model on the honeycomb lattice and detail
the three methods we apply.

\subsection{Model}
The Hamiltonian of the Heisenberg model is given by
\begin{subequations}
\begin{align}
    \mathcal{H} &= J \sum_{\langle i j \rangle} \vec{S_i} \vec{S_j} \\
    &= J \sum_{\langle i j \rangle} \left( S^z_i S^z_j + \frac{1}{2} \left( S^+_i S^-_j + S^-_i S^+_j \right) \right) \ ,
    \label{eq:hamiltonian}
\end{align}
\end{subequations}
where $S^{\pm,z}_i$ are spin operators acting on spins 1/2 arranged on a honeycomb lattice.
\begin{figure}
\centering
\includegraphics{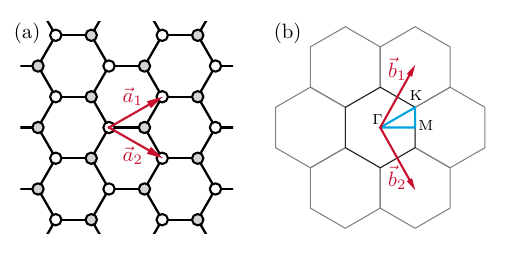}
\caption{Panel (a) shows a sketch of the honeycomb lattice and the elementary lattice vectors $\vec{a}_1 = (3/2, \sqrt{3}/2)$ and  \mbox{$\vec{a}_2 = (3/2, -\sqrt{3}/2)$}. In (b) a sketch of the Brillouin zone with reciprocal lattice vectors $\vec{b}_1 = (\pi, \sqrt{3} \pi)$, $\vec{b}_2 = (\pi, -\sqrt{3} \pi)$ and a high-symmetry path (blue) from the $\Gamma$-point $(0,0)$ to the $M$-point $(\pi,0)$ to the $K$-point $(\pi, \pi/\sqrt{3})$ is shown.}
\label{fig:honeycomb_sketch}
\end{figure}
We consider antiferromagnetic nearest-neighbor interactions with fixed $J$ as the energy unit, \ie, it is set to unity.
Since the honeycomb lattice is bipartite, there is no frustration in the system.
A sketch of the lattice and the corresponding Brillouin zone is shown in \cref{fig:honeycomb_sketch};
note that it is not a Bravais lattice, but has two sites per unit cell.
The ground state is long-range Néel ordered and breaks the SU(2) symmetry of the Hamiltonian \cite{Reger1989}.
The low-energy excitations can be understood as spin waves, \ie, quantized magnons
representing the gapless Goldstone bosons of the spontaneously broken continuous spin symmetry \cite{auerb94}.
In analogy to the Heisenberg model on the square lattice, the bipartiteness of the honeycomb Heisenberg model implies that magnons can only be created and annihilated in pairs. As a consequence, the sectors of odd and even magnon number are decoupled exactly by symmetry.

In momentum space one finds a one-magnon dispersion which is gapless at the $\Gamma$-points, but has a finite energy elsewhere.
The lower edge of multi-magnon continua is degenerate with the one-magnon dispersion.
This can be understood by the fact that the minimal multi-magnon energy at any total momentum $\vec{k}$ is given by a single magnon with momentum $\vec{k}$ and a certain number of gapless magnons at the $\Gamma$-point. This reasoning assumes that binding effects due to strong attractive magnon-magnon interactions 
are not relevant. If this is not true, binding can occur and  multi-magnon continua can be present at excitation energies below the expected one-magnon dispersion inducing  a quantum decay of magnons:
in the spectral response, sharp peaks from single magnons become broader multi-magnon resonances.

\subsection{Stochastic analytic continuation of quantum Monte Carlo data}

The dynamic structure factor encodes rich information about the excitation spectrum of a system and offers a direct bridge to inelastic neutron scattering experiments.
We can express this observable very generally as the response to an operator $\mathcal{O}_{\vec{k}}$
\begin{align}
    S_{\vec{k}}(\omega) \propto \sum_{n,m} e^{-\beta E_m} |\langle n| \mathcal{O}_{\vec{k}} | m \rangle |^{2} \delta\left(\omega-\left[E_n-E_m\right]\right) \, ,
\end{align}
where $|n\rangle$ are eigenstates of the Hamiltonian with corresponding energy $E_n$.
The dynamic structure factor $S_{\vec{k}}(\omega)$ depends on the transferred momentum $\vec{k}$ and the energy difference $\omega = E_n - E_m$.
In our quantum Monte Carlo (QMC) simulation we cannot directly access this quantity.
Instead we have to perform a numerical analytic continuation of correlation functions in imaginary time
\begin{align}
    G_{\vec{k}}(\tau) = \langle \mathcal{O}_{\vec{k}} (\tau) \mathcal{O}_{-\vec{k}}(0) \rangle
\end{align}
to obtain $S_{\vec{k}}(\omega)$.
There are several methods focusing on the analytic continuation of numerical data.
Most prominent are maximum entropy methods \cite{Gull1984, Jarrell1996, Silver1990} and the stochastic analytic continuation method \cite{Sandvik1998, Mishchenko2000, Vafayi2007, Beach2004, Shao2017, Shao2023}.
We consider the stochastic analytic continuation procedure as outlined in Ref.\,\cite{Shao2023}.
For the Heisenberg model on the honeycomb lattice we choose the operator $\mathcal{O}_{\vec{k}} = S_{\vec{k}}^z$, which is the Fourier transformed spin operator $S_i^z$.
Since we simulate finite systems the ground state does not spontaneously break the SU(2) symmetry.
The magnetic excitations are therefore strictly speaking "paramagnons" \cite{Doniach1966}, which are adiabatically connected to the magnon excitations in the thermodynamic limit.
In our finite-size simulation all spin components are equivalent and correlation functions are rotationally invariant.
Measuring $S_{\vec{k}}^z$ therefore yields the rotationally averaged dynamical structure factor, which can be viewed as an average over longitudinal and transverse correlations with respect to the Néel order parameter of the symmetry-broken state in the thermodynamic limit.
Technically it would be possible to separate the longitudinal and transversal contributions of the spectrum by introducing an additional staggered field (\eg in $z$-direction), that explicitly breaks the SU(2) symmetry, as done in Ref.\,\cite{Sandvik2001}.
But we do not follow this route here in order to avoid auxiliary fields such as the staggered field and its debatable value.
The QMC simulations are performed at large inverse temperatures $\beta = 1/T$.
As long as $T$ is smaller than the smallest energy gap in the finite system, the system effectively remains in its ground state, and the resulting dynamical structure factor corresponds to the zero-temperature limit
\begin{align}
    S_{\vec{k}}(\omega) \propto \sum_{n} |\langle n| \mathcal{O}_{\vec{k}} | 0 \rangle |^{2} \delta\left(\omega-\left[E_n-E_0\right]\right) \ .
    \label{eq:dsf}
\end{align}
In the following, we will first introduce the quantum Monte Carlo method and how imaginary-time correlation functions are measured during the simulation.
Then the stochastic analytic continuation method is presented.

\subsubsection{Stochastic series expansion quantum Monte Carlo}

We determine $G_{\vec{k}}(\tau)$ via the stochastic series expansion (SSE) quantum Monte Carlo method \cite{Sandvik1991, Sandvik1992, Sandvik2002, Sandvik2010}.
We follow the scheme described in Ref.\,\cite{Sandvik2010} for the isotropic Heisenberg model.
The Hamiltonian in \cref{eq:hamiltonian} can be rewritten in the language of lattice bonds $b$
\begin{align}
    \mathcal{H} = - \sum_b \left( \mathcal{H}_{1,b} - \mathcal{H}_{2,b} \right)
    \label{eq:sse-decomposition}
\end{align}
with the operators
\begin{subequations}
\begin{align}
    &\mathcal{H}_{1,b} = \frac{1}{4} - S^z_{i(b)} S^z_{j(b)} \\
    &\mathcal{H}_{2,b} = \frac{1}{2} \left( S^+_{i(b)} S^-_{j(b)} + S^-_{i(b)} S^+_{j(b)} \right) \ .
\end{align}
\end{subequations}
The SSE idea is to expand the partition function $Z$ in powers of $\beta \mathcal{H}$ and to utilize the decomposition of the Hamiltonian from \cref{eq:sse-decomposition}
\begin{subequations}
\begin{align}
    Z &= \mathrm{Tr}(e^{-\beta \mathcal{H}}) \\
	&= \sum_{ \{ | \alpha \rangle  \} } \sum_{n = 0}^\infty \frac{\beta^n}{n!} \, \langle \alpha |\, \Bigl( \ \sum_b \left( \mathcal{H}_{1,b} - \mathcal{H}_{2,b} \right) \Bigr)^n \, | \alpha \rangle \\
	&= \sum_{ \{ | \alpha \rangle  \} } \sum_{n = 0}^\infty \sum_{\{ S_n \}} \frac{\beta^n}{n!} \, \langle \alpha | \, \prod_{l = 1}^n \mathcal{H}_{a(l), b(l)} \, | \alpha \rangle \\
    &= \sum_{\omega \in \Omega} \pi(\beta, \omega)\ .
\end{align}
\end{subequations}
Written like this, $Z$ consists of all possible sequences $S_n$, which are the product of $n$ operators $\mathcal{H}_{a(l), b(l)}$, where $a(l)$ takes the values 1 or 2 and $b(l)$ is the bond acted on at position $l$ of the sequence.
An initial state $| \alpha \rangle$ is propagated through the sequence by acting on it with $\mathcal{H}_{a(l), b(l)}$.
Due to the periodicity of the trace, the final state after acting with $n$ operators must be $| \alpha \rangle$ again.
With $\{ | \alpha \rangle  \}$ chosen to be the $z$-basis and the operators $\mathcal{H}_{a(l), b(l)}$ defined above, no superpositions or negative amplitudes can be generated.
Note that we omitted the minus sign in front of $\mathcal{H}_{2,b}$ since there has to be always an even number of $\mathcal{H}_{2,b}$ operators in every sequence $S_n$ to satisfy the periodicity condition of the trace.
This does not hold for non-bipartite (geometrically frustrated) lattices, where this creates a sign problem.
We can now assign every object $\omega$ in the configuration space $\Omega =  \{ |\alpha \rangle \} \times \{ S_\mathcal{L} \}$ a weight $\pi(\beta, \omega)$, according to which we can do a Monte Carlo sampling.
The sum over $n$ is truncated at a fixed value $\mathcal{L}$, which can be determined during the equilibration such that only an exponentially small truncation error occurs \cite{Sandvik2002}.
All sequences with $n<\mathcal{L}$ are filled with identity operators $\mathds{1}$ and multiplied with a combinatorial factor to compensate for over-counting \cite{Sandvik2002}.
The configuration space $\Omega$ is updated through a diagonal update that inserts and removes diagonal operators $\mathcal{H}_{1, b(l)}$ based on the Metropolis Hastings algorithm and an off-diagonal cluster update that exchanges $\mathcal{H}_{1, b(l)}$ with $\mathcal{H}_{2, b(l)}$ and vice versa.
Using the beta-doubling technique \cite{Sandvik2002} we cool the system to sample effectively at $T=0$.
We refer to Refs.\,\cite{Sandvik2010, Adelhardt2024} for details on the updates and implementation.
\\ \\
Our goal is to compute the imaginary-time correlation function
\begin{align}
    G_{\vec{k}}(\tau) = \langle S_{\vec{k}}^z(\tau) S_{-\vec{k}}^z(0) \rangle \ ,
\end{align}
where $S_{\vec{k}}^z(\tau)$ is the Fourier-transformed spin state at imaginary time $\tau$
\begin{align}
    S_{\vec{k}}^z(\tau) = \sum_{j} e^{\tau \mathcal{H}} S^z_j e^{-\tau \mathcal{H}} e^{i \vec{k} \cdot \vec{r}_j} \ .
\end{align}
Comparing the SSE sequence to a continuous time formulation, imaginary times $\tau$ can be connected to the position $l$ of an operator in the sequence $S_\mathcal{L}$ \cite{Sandvik1992, Dorneich2001, Michel2007}.
This allows to assign an imaginary time to every operator and either measure $S_i(\tau)$ directly or transform the observable to the space of Matsubara frequencies \cite{Lohoefer2015, Lohoefer_Phd, Humeniuk_Phd}.
The latter is frequently used in the literature, and we have also implemented this method to calculate observables.
However, it turned out that the following method offers greater accuracy with less computational effort.
We choose the approach mentioned in Ref.\,\cite{Sandvik2019}, where the imaginary time axis is cut into $m$ regular time-slices of size $\Delta_\tau = \beta/m$.
This requires neither a probabilistic assignment of imaginary times to operators in the sequence nor a binning of measured values.
Here, the SSE approach outlined above is altered by expanding every slice $e^{-\Delta_\tau \mathcal{H}}$ individually.
The partition function is then given by
\begin{subequations}
\begin{align}
    Z &= \mathrm{Tr}(e^{-\beta \mathcal{H}}) = \mathrm{Tr}\Bigl( \bigl( e^{-\Delta_\tau \mathcal{H}} \bigr)^m \Bigr) \\
    &= \sum_{ \{ | \alpha \rangle  \} } \langle \alpha | \prod_{s=1}^m \Bigl( \sum_{n_s = 0}^\infty \sum_{\{ S_{n_s} \}} \frac{\Delta_\tau^{n_s}}{{n_s}!} \, \prod_{l = 1}^{n_s} \mathcal{H}_{a(l), b(l)} \, \Bigr) \, | \alpha \rangle \ .
\end{align}
\end{subequations}
In each time-slice $s$ the diagonal updates are carried out independently of the other slices with an individual number of operators per time-slice $n_s$.
The off-diagonal update is still carried out over the whole sequence since it does not alter the weight $\pi(\beta, \omega)$ of a configuration and preserves the periodicity of the trace.
The state between the slice $s$ and $s+1$ is then assigned to the imaginary time $\tau = s \cdot \Delta_\tau$ and each of these states is constantly measured during the simulation.

\subsubsection{Stochastic analytic continuation}
\label{sec:SAC}

The analytic continuation of imaginary-time Green's functions $G_{\vec{k}}(\tau)$ to spectral functions $S_{\vec{k}}(\omega)$ on the real frequency axis is a numerically ill-posed problem.
The stochastic analytic continuation (SAC) \cite{Sandvik1998, Mishchenko2000, Vafayi2007, Beach2004, Shao2017, Shao2023} utilizes a Monte Carlo sampling of the spectral function.
We refer to Ref.\,\cite{Shao2023} for an in-depth introduction and comparison of the method with other approaches like the Maximum Entropy method.
Here, we only want to highlight the most important concepts.
The SAC is carried out independently for each momentum, we therefore omit the index $\vec{k}$ in the following.
We start with a general ansatz for the spectral function as a sum of delta functions
\begin{align}
    S(\omega) = \sum_{i} S_i \, \delta(\omega-\omega_i)
    \label{eq:Skw_ansatz}
\end{align}
normalized by $\sum_i S_i = 1$.
The corresponding Green's function on the imaginary-time axis can be calculated by
\begin{align}
    G^S(\tau) = \int_{0}^\infty d\omega \ S(\omega) \, \frac{e^{-\tau \omega} + e^{-(\beta-\tau) \omega}}{1+e^{-\beta\omega}} \ .
\end{align}
We can now compare this with the Green's function $G(\tau)$ we determined using the SSE simulation.
$G(\tau)$ is normalized to match the definition of $S(\omega)$ in \cref{eq:Skw_ansatz}.
The correct amplitude is recovered after the SAC.
The goodness $\chi^2$ of the proposed $S(\omega)$ is given by
\begin{align}
    \chi^2(S(\omega), G(\tau)) = \sum_i^{N_\tau} \left( \frac{G^S(\tau_i) - G(\tau_i)}{\sigma(\tau_i)} \right)^2 \ ,
\end{align}
where $\sigma(\tau_i)$ is the statistical error of $G(\tau)$ at $\tau = \tau_i$ \footnote{Actually we use the information of the whole covariance matrix and transform $G(\tau)$ and $G^S(\tau)$ to the eigenbasis of the covariance matrix. We refer to Ref.\,\cite{Shao2023} for further details.} .
$S(\omega)$ can now be updated using the Metropolis-Hastings algorithm according to a Boltzmann-like probability distribution
\begin{align}
    P(S(\omega) | G(\tau) ) = \exp \left( -\frac{\chi^2(S(\omega), G(\tau))}{2 \theta} \right) \ ,
\end{align}
where $\theta$ is an artificial sampling temperature.
There are several techniques to determine the optimal sampling temperature (see \eg Refs.\,\cite{Shao2023, Beach2004}).
In this work, we use the criterion outlined in Ref.\,\cite{Shao2023}.
Our results, however, were quite stable under variation of $\theta$.
Generally one can use updates that modify the amplitudes $S_i$, the positions $\omega_i$ of the delta functions or both.
The differences in the results are rather minor \cite{Shao2023}.
However, if one already has information about the form of the spectrum, one can use this to constrain the simulation.
In the case of the antiferromagnetic Heisenberg model on the honeycomb lattice we expect a sharp quasi-particle peak (magnon-peak) followed by a continuum of higher excitations.
\begin{figure}
\centering
\includegraphics{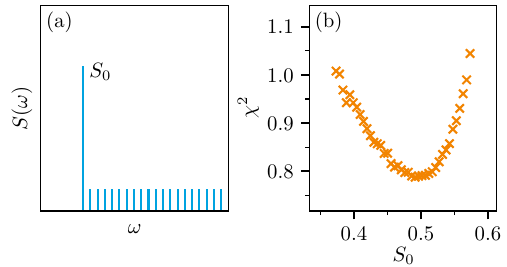}
\caption{In the restricted sampling procedure the ansatz for $S(\omega)$ contains a delta peak with fixed amplitude $S_0$ followed by a continuum of equal amplitude delta functions. The ansatz for $S(\omega)$ is schematically shown in panel (a). Comparing $\chi^2$ for different simulations at fixed values $S_0$, the optimal $S_0$ can be determined. In panel (b), $\chi^2(S_0)$ for the $L=36$ Heisenberg model on the honeycomb lattice at the $M$-point is shown.}
\label{fig:Sw_ansatz}
\end{figure}
Therefore, we constrain the spectrum in the following way:
We fix the amplitude $S_0$ of the first delta peak and distribute the remaining weight $1-S_0$ equally to the remaining delta functions (see \cref{fig:Sw_ansatz}a).
Then we update only the positions of the delta functions (including $\omega_0$) with the constraint $\omega_i > \omega_0 \ \forall i >0$.
Comparing the optimal $\chi^2$ for different fixed $S_0$ allows to determine the optimal $S_0$ (see \cref{fig:Sw_ansatz}b).
This has proven to be successful for studying $S(\omega)$ of the antiferromagnetic 
Heisenberg model and $J$-$Q$-model on the square lattice \cite{Shao2017} and 
additionally allows us to determine the weight on the one-magnon peak.

\subsubsection{Simulation details}

All SSE QMC calculations are carried out at inverse temperature $\beta=256$ with $s=2560$ 
steps in imaginary time resulting in a resolution of $\Delta_\tau = 0.1$.
This temperature is sufficient to simulate $T=0$ observables throughout the momentum space apart from the gapless $\Gamma$ points.
We consider systems up to linear size $L=36$, which equals 2592 spins, with ${(L\bmod 6\stackrel{!}{=}0)}$ such that the $\vec{k}$ vectors of interest $2\vec{b}_1/3 + \vec{b}_2/3$ ($K$-point) and $\vec{b}_1/2 + \vec{b}_2/2$ ($M$-point) lie on the reciprocal lattice.
For the SAC we take into account $N_\tau \sim L$ points in imaginary time with a linear grid of $\tau_i$ where the relative error satisfies $\sigma(\tau_i)/G_{\vec{k}}(\tau_i) < 0.2$.
Using unrestricted SAC sampling we first determine the minimal $\chi^2_{\mathrm{min}}$ and then the optimal sampling temperature from the condition $\theta_\mathrm{opt} = \theta(\chi^2=\chi^2_{\mathrm{min}} +a)$ \cite{Shao2023}.
Varying $a$ did not significantly vary our results.
We set $a=1$ in this work unless otherwise stated.
After fixing $\theta_\mathrm{opt}$ all following simulations are performed at the same $\theta=\theta_\mathrm{opt}$.

\subsection{Continuous similarity transformations}

We utilize the approach outlined in previous studies \cite{Powalski2015,Powalski2018,Walther2023,Hering2024}.
First, a self-consistent mean-field solution in the thermodynamic limit is 
determined for the Hamiltonian of \cref{eq:hamiltonian} represented by
Dyson-Maleev bosons. Subsequently, we solve the resulting Hamiltonian 
with a \gls{cst}, allowing us to describe the gapless Goldstone bosons associated with the N\'eel phase.

Regarding the first step, we start with the classical N\'eel state. 
Applying the Dyson-Maleev transformation \cite{Dyson_1956,malee58b}, we obtain a bosonic Hamiltonian that is inherently non-Hermitian.
Successively, a mean-field decoupling \cite{Takahashi_1989,Hida_1990}, a Fourier transformation, and a Bogoliubov transformation are carried out for this Hamiltonian. 
The Bogoliubov transformation is implemented in a self-consistent fashion, diagonalizing the bilinear part of the Hamiltonian.

Due to the non-Hermitian nature of the Dyson-Maleev Hamiltonian, we employ a \gls{cst} rather than the more commonly used \glsentrylong{cut} \cite{Wegner_1994,Mielke_1998,Knetter_2000,Knetter_2003}.
Despite this difference, the underlying concept remains the same.
The initial quantum many-body Hamiltonian $\mathcal{H}_0$ is continuously transformed into a simpler, more manageable effective Hamiltonian $\mathcal{H}_\text{eff}$.
This transformation can be induced by a given generator $\eta(\ell)$, and the corresponding flow equation reads
\begin{align} 
\partial_{\ell} \mathcal{H}(\ell) = \left[ \eta(\ell), \mathcal{H}(\ell) \right]~. \label{eq:flow_equation}
\end{align}
In the limit $\ell\to\infty$ we obtain the effective Hamiltonian ${\mathcal{H}(\ell=\infty)=\mathcal{H}_\text{eff}}$, which is characterized primarily by its simpler form, \eg, a (block-)diagonal form. 
The specific choice of the generator $\eta(\ell)$ affects this final form.

In this work, we use the \gls{qpc} generator and the $0n$ generator.
The \gls{qpc} generator contains the same terms as the Hamiltonian with the same sign if they
increase the number of bosons and the opposite one if they decrease it. 
If they do not change this number these terms are omitted. 
The $0n$ generator only keeps terms of the \gls{qpc} 
generator which either consist only of creation or only of annihilation operators. 
Generally, one can consider also a $1n$ generator which contains the terms of the $0n$ plus
those which contain either one annihilation operator and two and more creation operators or one creation operator and two and more annihilation operators.

Subsequently, the main features of both generators are discussed.
The \gls{qpc} generator~\cite{Knetter_2000, Mielke_1998, Fischer_2010} separates all quasi-particle sectors ensuring that for a convergent flow, \ie, the flow equation \cref{eq:flow_equation} converges to a fixed point $\mathcal{H}_{\text{eff}}$, only operators occur in $\mathcal{H}_{\text{eff}}$ that conserve the number of quasi-particles, in this case, magnons.
However, a convergence of the flow with this generator can only be reached if there are no energetic overlaps between different quasi-particle sectors.
For example, if binding effects in the three-quasi-particle (3QP) sector lead to an energy that is lower than that of the single-quasi-particle (1QP) sector at the same momentum, the flow diverges and the quasi-particle decays.

While the \gls{qpc} generator disentangles all magnon sectors from one another,
the $1n$ generator disentangles only the ground state and the one-magnon states from the
remaining Hilbert space. But since we will restrict our analysis to quartic terms in 
second quantization and magnon number only changes in pairs the $1n$ and the \gls{qpc}
generator coincide.

The $0n$ generator \cite{Fischer_2010} is less ambitious, so that for a convergent flow, 
only the ground state is disentangled from all higher quasi-particle sectors in 
$\mathcal{H}_{\text{eff}}$.
This comes with the advantage that the flow of the $0n$ generator is generally more robust, 
since energetic overlaps between higher quasi-particle sectors are irrelevant for the convergence behavior. 
Only an overlap of the ground-state energy (0QP) sector with the multi-quasi-particle sector
($\geq$1QP) results in a diverging flow.
Therefore, the $0n$ generator can also be used to investigate phase transitions out of ordered phases by analyzing the stability of the flow~\cite{Hering2024}. 

Besides the choice of the generator, computing the flow equations additionally requires a suitable truncation scheme. 
This is due to the fact that the evaluation of $\left[ \eta(\ell), \mathcal{H}(\ell) \right]$ generally results in infinitely many arbitrarily complicated operators. 
Here, we use the truncation scheme based on the scaling dimension $\dsc$ \cite{Powalski2015, Powalski2018, Walther2023, Caci2024, Hering2024}.
This enables us to systematically take into account the most relevant magnon-magnon interaction with the \gls{cst}.
Operators with a higher scaling dimension, especially for gapless phases, are of lesser significance in comparison to operators with a low scaling dimension. 
The \gls{cst}, using the \gls{qpc} generator together with a truncation of operators with $\dsc>2$, was successfully applied to determine the one-magnon dispersion and the dynamical structure factor of the magnetically ordered phase in the antiferromagnetic spin-$1/2$ Heisenberg model on the square lattice~\cite{Powalski2015,Powalski2018}.
As a natural extension, the single- and two-particle properties of the easy-axis spin-$1/2$ XXZ model on the square lattice could be reproduced quantitatively~\cite{Walther2023,Caci2024}.
Applying the \gls{cst} with the $0n$ generator, critical points of phase transition between magnetically ordered phases and nonmagnetic phases were determined for the Heisenberg bilayer and the $J_1$-$J_2$ model~\cite{Hering2024}.
So far, all studies referred to dealt with variants of the square lattice.
Consequently, adapting in this work the \gls{cst} to the antiferromagnetic spin-$1/2$ 
Heisenberg model on the honeycomb lattice promises to be an intriguing and fruitful endeavor.

\subsubsection{Implementation details}

In terms of the numerical implementation of the flow equations,
solving themin momentum space requires a discretization of the Brillouin zone. 
Here, periodic boundary conditions are used that contain the $\Gamma$-point.
The $\Gamma$-point is numerically challenging because 
the mean-field solution displays an integrable divergence at this point.
To circumvent this numerical problem,
 we set coefficients containing the $\Gamma$-point to zero, 
 effectively disabling them during the flow.
This was already done in Refs.~\cite{Powalski2015,Powalski2018,Hering2024}.
Using different linear system sizes $L$ in the discretization enables
us to  extrapolate the results to the thermodynamic limit.
Additionally, since we are explicitly interested in the physics of the $K$-point,
we only consider system sizes $L$ for which the $K$-point  
belongs to the discretization grid ${(L\bmod 6\stackrel{!}{=}0)}$, 
with the largest system size of $L=18$.

Furthermore, the limit $\ell\to\infty$ is numerically unfeasible;
instead the flow is stopped if the \gls{rod}, \ie, the square root of the sum over all squared entries of the generator $\eta(\ell)$, falls below a threshold value of $10^{-6}J$. 
Since this threshold is independent of the linear system size $L$ and  the number of entries grows 
$\propto L^6$, the numerical accuracy increases for higher $L$
because the residual terms must be reduced even further.

\subsubsection{Convergence}
\label{sssec:convergence}

In contrast to the converging flow of the \gls{qpc} generator for the antiferromagnetic Heisenberg model on a square lattice, we observe divergent flows for the honeycomb lattice for all considered system sizes up to $L=18$.
Therefore, it is not possible to reach an effective Hamiltonian that conserves the number of particles with the  \gls{cst} approach.
We interpret a diverging flow as a signature 
of a relevant overlap of the one-magnon and the three-magnon subspaces which in turn
implies the decay of the one-magnon states so that 
the weight of the single magnon peak in the spectral response vanishes 
at least somewhere in the Brillouin zone.
Due to such a decay channel, the \gls{qpc} generator is unable to fully order the 
 sectors of different magnon number energetically which results in a divergent flow. 
In the following, we discuss two modifications of the approach, 
allowing us to gain further insight in the honeycomb Heisenberg model by \gls{cst}:

First, we apply the $0n$ generator which only aims to decouple the ground state from all excited states.
We find a convergent flow for all considered system sizes confirming that the long-range ordered
state is indeed the ground state as expected. A diverging flow 
would have implied that another state is lower in energy so that the assumed ground state
was the wrong state.
Although the flow converges with the $0n$ generator, the single-magnon sector remains 
connected to the three-magnon sector in the effective Hamiltonian. 
Consequently, it is reasonable to consider the one-three magnon subspace for each $\vec{k}$ point 
and compute the lowest eigenvalue as estimate for the single-magnon dispersion $\tilde{\omega}_{0n}$.
Note that one should also consider 5 and 7 and so on magnon states. But this exceeds the numerical
possibilities and would lead only to minor corrections.

The second approach uses the \gls{qpc} generator and, instead of expecting convergence of the flow, takes advantage of the fact that it is also possible to stop the flow at any value of the flow parameter 
$\ell$. As long as the \gls{rod} is decreasing, we expect that the rotation to an effective model is not driven by diverging non-physical terms. 
Of course, if the flow is stopped at a certain finite $\ell$, off-diagonal terms still remain. 
Nevertheless, the effective Hamiltonian can be expected to contain more relevant interactions, stemming from rotating away some of the initial off-diagonal terms.
A plausible point to stop the flow is where the \gls{rod} reaches a minimum before it begins to diverge.
However, we find that a subdivision of the \gls{rod} into all different types of terms is reasonable.
We consider the $n:m$ subdivisions consisting of the terms with $n$ annihilation operators and $m$
creation operators and vice versa. Hence in our case, there are the subdivisions $0:2$, $0:4$ and $1:3$.
The most critical subdivision is $1:3$ because it comprises the terms implying quantum decay 
of a magnon into three-magnon states. Thus, we focus on the \gls{rod} of this subdivision
and search for its minimum for smaller values of $\ell$ compared to the minimum of the full \gls{rod}.
At this minimum, we stop the flow.
Interestingly, this minimum hardly shows any finite-size scaling and it is 
located close to $\ell=1/J$ for all system sizes.  
Thus, in this approach, the flow with the \gls{qpc} generator is stopped 
before the flow parameter exceeds $\ell>1/J$.
The resulting effective Hamiltonian still contains terms that do not conserve the number of magnons.
Hence, the dispersion $\tilde{\omega}_{\mathrm{qpc}}$ is obtained in the same fashion as for the $0n$ generator, \ie, by diagonalizing the one-three magnon subspace. 
An alternative option is to analyze only the bare dispersion 
after the stopped \gls{qpc} flow yielding $\omega_{\mathrm{qpc}}$  or after  the 
converged $0n$ flow $\omega_{0n}$ neglecting the remaining couplings to higher magnon spaces.

\subsection{Series expansions}

Series expansions (SE) allow for the exact determination of high-order series expansion of excitation energies in the thermodynamic limit about a perturbative limit, which is complementary to QMC and \gls{cst}. Specifically, we introduce the anisotropy parameter $\lambda$ in Eq.~\eqref{eq:hamiltonian} and investigate the XXZ-model
\begin{align}
    \mathcal{H}(\lambda) &= J \sum_{\langle i j \rangle} \left( S^z_i S^z_j + \lambda \, \frac{1}{2} \left( S^+_i S^-_j + S^-_i S^+_j \right) \right) 
\end{align}
so that $\mathcal{H}(\lambda=0)$ corresponds to the Ising limit and $\mathcal{H}(\lambda=1)$ to the Heisenberg point. Applying a sublattice rotation by $\pi/2$ about the $y$-axis on sublattice $B$ gives
\begin{align}
\tilde{\mathcal{H}}(\lambda ) &= -J \sum_{\langle i j \rangle} \left( S^z_i S^z_j + \lambda \, \frac{1}{2} \left( S^+_i S^+_j + S^-_i S^-_j \right) \right)\\ 
        &= \mathcal{H}_0 + \lambda V
    \label{eq:ham_xxz}
\end{align}
so that the unperturbed Hamiltonian at $\lambda=0$ corresponds to the ferromagnetic Ising model on the honeycomb lattice which has two symmetry-equivalent ferromagnetic ground states with all spins pointing either down or up. We choose one of the states $|\mathrm{ref}\rangle = |\uparrow \dots \uparrow\rangle$ as the reference state for the SE in the ordered phase with an unperturbed energy $E_0^{(0)} = -JN/2$. Quasi-particle (QP) excitations in the ordered ferromagnet at finite $\lambda$ correspond to dressed spin-flip excitations (magnons) adiabatically connected to local spin flips in the unperturbed limit. The unperturbed one-spin-flip states are therefore $|i\rangle_0 = S^+_i|\mathrm{ref}\rangle$.
Here we apply SE to calculate the 1QP dispersion up to order 14 in $\lambda$ in the thermodynamic limit. Pad\'{e} extrapolation \cite{Guttmann1989}
is then used to estimate the one-magnon excitation spectrum of the Heisenberg model on the honeycomb lattice.

To reach high orders, the honeycomb lattice is decomposed into finite linked graphs, and effective Hamiltonians are derived on these graphs which isolate the one-magnon processes. The projective-cluster additive transformation (PCAT) \cite{hormann2023projective} is employed, a general framework that enables the construction of cluster-additive effective Hamiltonians in the thermodynamic limit from finite-cluster calculations.
Within the PCAT framework, the two-block orthogonalization method (TBOT) \cite{oitmaa2006series} computes the effective 1QP Hamiltonians on individual clusters. TBOT is equivalent to a Schrieffer-Wolff transformation and depends only on eigenvectors of the degenerate subspace, offering the numerical advantage that only matrix-vector multiplications are required rather than full matrix diagonalizations.

\subsubsection{Cluster Hamiltonians and boundary fields}
Calculations are performed on induced linked subgraphs (linked clusters) $G$ of the honeycomb lattice. For each cluster $G$, the complement (environment) is denoted as $G^C$. The key assumption is that the total state factorizes as a product state between cluster and environment:
\begin{equation}
|\Psi\rangle_{\text{total}} = |\Psi\rangle_G \otimes |\mathrm{ref}\rangle_{G^C} ,
\end{equation}
where all environment spins remain frozen in the reference state $|\mathrm{ref}\rangle_{G^C} = |\uparrow\ldots\uparrow\rangle_{G^C}$. Under this assumption, the environment degrees of freedom can be traced out to obtain an effective cluster Hamiltonian $\tilde{\mathcal{H}}^{(G)}$ acting only on the cluster.

The cluster Hamiltonian consists of two parts:
\begin{equation}
\tilde{\mathcal{H}}^{(G)} = \tilde{\mathcal{H}}_{\text{bulk}}^{(G)} + \tilde{\mathcal{H}}_{\text{boundary}}^{(G)}
\end{equation}
where the bulk term contains all bonds within the cluster
\begin{equation}
\tilde{\mathcal{H}}_{\text{bulk}}^{(G)} = -J \sum_{\langle ij \rangle \in G} \left( S^z_i S^z_j + \lambda \frac{1}{2} (S^+_i S^+_j + S^-_i S^-_j) \right)
\end{equation}
and the boundary field term arises from bonds connecting cluster sites to the frozen environment
\begin{equation}
\begin{split}
\tilde{\mathcal{H}}_{\text{boundary}}^{(G)} = -J \sum_{\substack{i \in G, j \in G^C \\ \langle ij \rangle}} \Bigg( S^z_i \langle S^z_j \rangle \\
+ \frac{\lambda}{2} (S^+_i \langle S^+_j \rangle + S^-_i \langle S^-_j \rangle) \Bigg) .
\end{split}
\end{equation}
Since the environment spins are in the state $|\uparrow\rangle$, we have $\langle S^z_j \rangle = +1/2$, $\langle S^+_j \rangle = 0$, and $\langle S^-_j \rangle = 0$. Therefore:
\begin{equation}
\tilde{\mathcal{H}}_{\text{boundary}}^{(G)} = -\frac{J}{2} \sum_{\substack{i \in \partial G}} z_i S^z_i
\end{equation}
where $\partial G$ denotes the boundary sites of cluster $G$ and $z_i$ is the number of environment neighbors of site $i$. This boundary term acts therefore as an effective magnetic field on boundary sites.

With this construction, the unperturbed one-spin-flip states $|i\rangle_0$ form a degenerate subspace with energy \mbox{$E_1^{(0)} = E_0^{(0)} + 3J/2$}, independent of the position of the site $i$ in the cluster, because the boundary sites receive compensatory contributions from the boundary field.

\subsubsection{TBOT and PCAT Procedure}

Using TBOT, the perturbed eigenstates in the full $2^N$-dimensional Hilbert space are calculated for each cluster: (1) The ground-state eigenvector \mbox{$|\Psi^0\rangle = \sum_{k=0}^{n_{\max}} \lambda^k |\Psi^0\rangle^{(k)}$} up to order $n_{\max}$ in $\lambda$. (2) A basis of the one-spin-flip subspace $|\Psi^1_i\rangle = \sum_{k=0}^{n_{\max}} \lambda^k |\Psi^1_i\rangle^{(k)}$ for $i=1,\ldots,N$.
All quantities are expanded as formal power series in $\lambda$ and truncated at order $n_{\max}$. Each perturbation order is stored separately as a $2^N$-dimensional vector, and all subsequent operations must carefully track the perturbative order throughout.

To construct the PCAT transformation, the subtraction scheme removes ground-state admixtures from the excited states. The overlaps with the unperturbed ground state $|\Phi^0\rangle = |\mathrm{ref}\rangle$ can be directly read from the first component of the eigenvectors in the computational basis where $|\mathrm{ref}\rangle$ corresponds to the first basis vector:
\begin{itemize}
    \item $\langle\Phi^0|\Psi^0\rangle$ is the first component of $|\Psi^0\rangle$
    \item $\langle\Phi^0|\Psi^1_i\rangle$ is the first component of $|\Psi^1_i\rangle$
\end{itemize}

The modified 1QP states are constructed order by order
\begin{equation}
|\widetilde{\Psi}^1_i\rangle = |\Psi^1_i\rangle - \frac{\langle\Phi^0|\Psi^1_i\rangle}{\langle\Phi^0|\Psi^0\rangle}|\Psi^0\rangle .
\end{equation}
This subtraction ensures that $P^0|\widetilde{\Psi}^1_i\rangle = 0$, \ie, the modified states have no projection onto the unperturbed ground-state subspace.

Next, the modified states are projected onto the one-spin-flip subspace by extracting the $N$ components corresponding to the basis states $\{|i\rangle_0\}_{i=1}^N$. This yields an $N \times N$ overlap matrix $\widetilde{X}^1$, which is generally not orthonormal.

To obtain an orthonormal basis, symmetric orthogonalization 
\begin{equation}
\widetilde{X}^1_{\mathrm{ortho}} = \widetilde{X}^1 (\widetilde{X}^{1\dagger} \widetilde{X}^1)^{-1/2}
\end{equation}
is applied. Standard transformation schemes \cite{Cederbaum1989} construct the transformation directly from the unmodified overlap matrix $X^1$. However, because the perturbed one-spin-flip states have overlap with the unperturbed ground state, this approach lacks cluster additivity. Here, the orthogonalization is applied to the modified overlap matrix $\widetilde{X}^1$ instead, which ensures cluster additivity. The inverse square root is expanded order by order in $\lambda$ to maintain the perturbative structure.

Finally, the effective 1QP Hamiltonian is computed in this orthogonalized basis. The PCAT transformation for the 1QP subspace is
\begin{equation}
U_{\mathrm{PCAT}}^{\rm 1QP} = U^{\rm 1QP} \widetilde{X}^{1\dagger} (\widetilde{X}^1 \widetilde{X}^{1\dagger})^{-1/2} ,
\end{equation}
where $U^{\rm 1QP}$ has the perturbed 1QP eigenstates $|\Psi^1_i\rangle$ as columns. The effective Hamiltonian in the 1QP subspace is then
\begin{equation}
H_{\mathrm{eff}}^{\rm 1QP} = (U_{\mathrm{PCAT}}^{\rm 1QP})^\dagger \tilde{\mathcal{H}}^{(G)} U_{\mathrm{PCAT}}^{\rm 1QP} - E_0 \mathds{1} ,
\end{equation}
where the ground-state energy $E_0$ is subtracted from the diagonal. This subtraction is necessary to ensure additivity on disconnected clusters in the linked cluster expansion (LCE) \cite{Gelfand1996}.

\subsubsection{Thermodynamic limit}

The effective 1QP Hamiltonian in the thermodynamic limit is constructed using the LCE with an inclusion-exclusion principle. For each cluster $G$, the reduced effective cluster Hamiltonian $\bar{H}_{\mathrm{eff}}^{\rm 1QP}(G)$ is computed recursively by subtracting contributions from all proper subclusters
\begin{equation}
\bar{H}_{\mathrm{eff}}^{\rm 1QP}(G) = H_{\mathrm{eff}}^{\rm 1QP}(G) - \sum_{G' \subset G} \bar{H}_{\mathrm{eff}}^{\rm 1QP}(G') .
\end{equation}

The reduced contributions from all clusters are then embedded back onto the infinite honeycomb lattice. For each unique cluster topology G, the sum runs over all inequivalent embeddings on the honeycomb lattice
\begin{equation}
H_{\mathrm{eff}}^{\rm 1QP,\mathrm{lattice}} = \sum_{G} \sum_{\text{embeddings}} \bar{H}_{\mathrm{eff}}^{\rm 1QP}(G) .
\end{equation}
Two embeddings are considered equivalent if they are related by a lattice translation, which reduces the infinite set of all possible embeddings to a finite set of inequivalent ones. Order $n = 14$ is reached in the perturbative expansion, with the largest clusters containing up to $n+1 = 15$ bonds.

\subsubsection{Effective One-Magnon Hamiltonian}

The resulting effective one-magnon Hamiltonian on the lattice takes the form
\begin{equation}
\begin{split}
H_{\mathrm{eff}}^{\rm 1QP,\mathrm{lattice}} = \sum_{\vec{R},\alpha} E_1^{\alpha}(\lambda) |\vec{R},\alpha\rangle\langle\vec{R},\alpha| \\
+ \sum_{\vec{R},\vec{R}',\alpha,\beta} t_{\alpha\beta}(\vec{R}-\vec{R}';\lambda) |\vec{R},\alpha\rangle\langle\vec{R}',\beta| ,
\end{split}
\end{equation}
where $\vec{R}$ labels the unit cell, $\alpha, \beta \in \{A,B\}$ label the sublattices, and $E_1^{\alpha}(\lambda)$ and $t_{\alpha\beta}(\vec{\delta};\lambda)$ are power series in $\lambda$ 
\begin{align}
E_1^{\alpha}(\lambda) &= \sum_{n=0}^{n_{\max}} E_1^{\alpha,(n)} \lambda^n \\
t_{\alpha\beta}(\vec{\delta};\lambda) &= \sum_{n=0}^{n_{\max}} t_{\alpha\beta}^{(n)}(\vec{\delta}) \lambda^n
\end{align}
obtained from the PCAT procedure.

\subsubsection{1QP Dispersion}

The 1QP dispersion relation is obtained by Fourier transforming the effective 1QP Hamiltonian. For each momentum $\vec{k}$ one obtains the $2 \times 2$ matrix
\begin{equation}
H_{\mathrm{eff}}(\vec{k}) = \begin{pmatrix} E_1^A + \sum_{\vec{\delta}} t_{AA}(\vec{\delta}) e^{{\rm i}\vec{k}\cdot\vec{\delta}} & \sum_{\vec{\delta}} t_{AB}(\vec{\delta}) e^{{\rm i}\vec{k}\cdot\vec{\delta}} \\ \sum_{\vec{\delta}} t_{BA}(\vec{\delta}) e^{{\rm i}\vec{k}\cdot\vec{\delta}} & E_1^B + \sum_{\vec{\delta}} t_{BB}(\vec{\delta}) e^{i\vec{k}\cdot\vec{\delta}} \end{pmatrix}
\end{equation}
reflecting the two-site unit cell of the honeycomb lattice. For the XXZ model on the bipartite honeycomb lattice all hopping elements $t_{AB}$ and $t_{BA}$ vanish. In addition, one has $t_{AA}=t_{BB}$ which have only contributions in even order in $\lambda$. The two degenerate eigenvalues correspond to the 1QP dispersion $\omega(\vec{k})$.

\subsubsection{Pad\'{e} extrapolation}
\label{sssec:pade}
In order to estimate the 1QP dispersion $\omega(\vec{k})$ at the Heisenberg point $\lambda=1$ we use Pad\'{e} extrapolation which is a 
standard method to extrapolate the series beyond their original radius of convergence \cite{Guttmann1989}. 
Because the series contain only even-order contributions in $\lambda$, we introduce 
$x \coloneqq \lambda^2$ to obtain $\omega(\vec{k})$ as a series in $x$ up to order 7. The Pad\'e approximant $G^{L/M}(x)$ is then defined as a fraction of two polynomials that give the series of $\omega(\vec{k})$ as a Taylor expansion in order $7$ about $x=0$
\begin{align}
    G^{L/M}(x)=\frac{p_0+p_1 x+p_2 x^2+...+p_L x^L}{q_0+q_1 x+q_2 x^2+...+q_M x^M} \ ,
\end{align}
where degree of the denominator (numerator) is denoted by $M$ ($L$).
The definition via the Taylor expansion provides a sufficient set of linear equations that can be solved for given parameters as long as $L+M\leq 7$.

In practice, we consider the four Pad\'e approximants $G^{5/2}(x)$, $G^{4/3}(x)$, $G^{3/4}(x)$, and $G^{2/5}(x)$. The mean value and its standard deviation at $\lambda=1$ serve as an estimate for and uncertainty of the 1QP dispersion of the Heisenberg model on the honeycomb lattice. 
The chosen extrapolation scheme is not expected to work well for two cases. First, when the gap closes with a power-law behaviour which is expected for $\vec{k}=\Gamma $ with $(1-\lambda)^{1/2}$ close to the Heisenberg point. This power-law behaviour can in principle be described by Dlog Pad\'{e} extrapolation. However, as already observed in Ref.~\cite{Kadosawa2024}, the gap series is very hard to extrapolate due to strongly increasing coefficients of higher-order terms. Second, the decay of quasi-particles corresponds to the breakdown of the quasi-particle picture. Here we expect to observe large uncertainties in the Pad\'{e} extrapolation. 

\section{Results}
\label{sec:results}

\begin{figure*}
    \centering
    \includegraphics{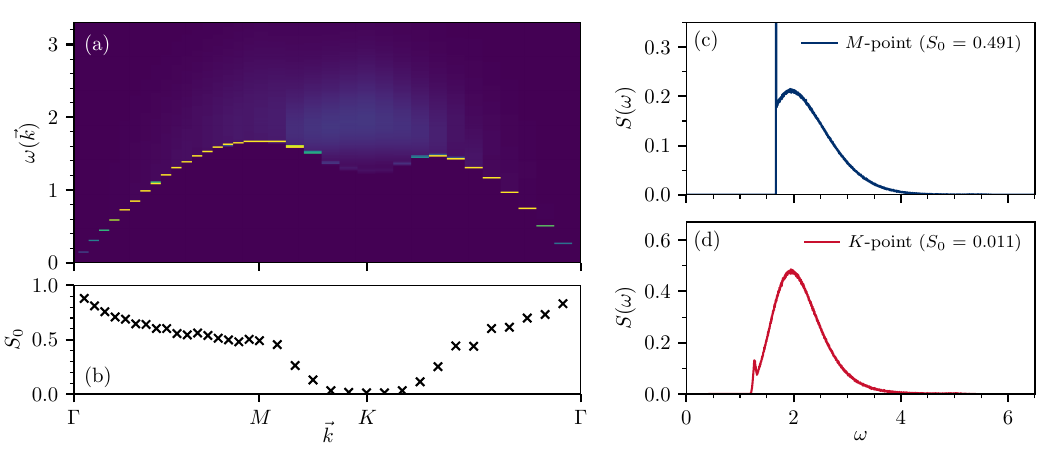}
    \caption{The dynamic structure factor $S_{\vec{k}}(\omega)$ of the Heisenberg antiferromagnet on a $L=36$ honeycomb lattice on the high-symmetry path $\Gamma \rightarrow M \rightarrow K \rightarrow \Gamma$ (see \cref{fig:honeycomb_sketch}) is shown in panel (a). The spectrum is calculated using the restricted SAC sampling scheme. The corresponding optimal amplitudes $S_0$ of the first delta peak in the simulation are depicted in panel (b) below. In (c) and (d), cuts through the dynamic structure factor at the M- and K-point with their optimal $S_0$ are shown.}
    \label{fig:Skw}
\end{figure*}

In this section we present results for the antiferromagnetic Heisenberg model on the honeycomb lattice.
We begin with the dynamic structure factor obtained from QMC simulations combined with SAC.
\cref{fig:Skw}a shows the dynamic structure factor $S_{\vec{k}}(\omega)$ for a system of size $L=36$ along the high-symmetry path $\Gamma \rightarrow M \rightarrow K \rightarrow \Gamma$ in the Brillouin zone, which is illustrated in \cref{fig:honeycomb_sketch}b. 
For each momentum, we apply the restricted SAC sampling scheme introduced in \cref{sec:SAC} to determine the optimal weight $S_0$ of the lowest-energy delta peak and the corresponding spectrum.
Along most of the high-symmetry path the spectrum at low energies is dominated by a pronounced macroscopic peak corresponding to the one-magnon mode. 
This peak disappears at the $K$-point, where the spectral weight is instead distributed over a broad continuum extending over a wide energy range.
The resulting spectra agree well with the experimental and numerical results of the honeycomb Heisenberg antiferromagnet in Ref.\,\cite{Hernandez2025}.
Qualitatively the dynamic structure factors in Refs.\,\cite{Ferrari2020, Gu2022} are similar to our results, but differ in the distribution of spectral weight, which likely can be traced back to approximations taken in their approaches.
Looking at the lower band edge of the spectrum we observe a pronounced minimum analogous to the ``roton minimum'' found in the square-lattice Heisenberg model at $\vec{k}=(\pi,0)$ \cite{Sandvik2001, Powalski2015, Shao2017, Powalski2018}.
The optimal weight $S_0$ of the lowest-energy delta peak is shown in \cref{fig:Skw}b.
While $S_0$ remains finite over most of the Brillouin zone, it drops sharply when approaching the $K$-point and becomes nearly zero there. 
This behavior is further highlighted by the momentum-resolved cuts of $S_{\vec{k}}(\omega)$ shown in panels (c) and (d) of \cref{fig:Skw}, where the contrast between the spectra at the $M$- and $K$-points is particularly evident.
Compared to the square-lattice Heisenberg model \cite{Shao2017}, the overall structure of the spectra is similar, but the suppression of the one-magnon weight at the $K$-point is much more pronounced, with almost all spectral weight residing in the continuum.

\begin{figure}
\centering
\includegraphics{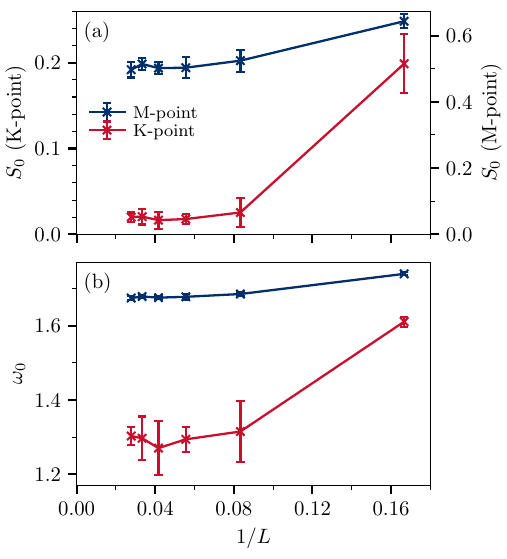}
\caption{Scaling of the optimal $S_0$ (a) and corresponding position of the first delta peak $\omega_0$ (b) with different system sizes $L$ at the $K$- and $M$-point. The simulation and optimization is repeated for different sampling temperatures $\theta$ and independent sets of QMC data. The displayed error bars represent the standard deviation calculated across the set of independent simulation runs.}
\label{fig:scaling_S0}
\end{figure}

Within the restricted SAC approach it is difficult to distinguish rigorously between strictly zero weight and a very small but finite weight of the lowest-energy peak. 
To clarify this issue, we analyze the system-size dependence of $S_0$ at the $K$- and $M$-point by repeating the simulations for several lattice sizes $L$.
The results are depicted in \cref{fig:scaling_S0}.
\cref{fig:scaling_S0} shows the optimal values of $S_0$ and the corresponding peak position $\omega_0$ as functions of the system size $L$.
To obtain reasonable error estimates, we generated five independent sets of QMC imaginary-time correlation functions and, for each set, repeated the SAC optimization for five different sampling temperatures $\theta$ using parameters $a \in [0.5,3]$.

At the $M$-point the one-magnon weight converges to a finite value of approximately $S_0 \approx 0.5$.
In contrast, at the $K$-point $S_0$ shifts towards zero.
The weight does not reach exactly zero, since it is strictly positive by construction.
The data points of the largest system sizes deviate only within the statistical accuracy from one another.
The reduction of $S_0$ is accompanied by a downward shift of the lowest excitation energy $\omega_0$.
This redistribution of spectral weight from the one-magnon peak to the continuum and the associated local minimum in the dispersion are also observed in the square-lattice Heisenberg model \cite{Shao2017}.
The absence of a finite delta peak at the lower edge of the spectrum at the $K$-point indicates that, unlike at other momenta, no well-defined one-magnon excitation survives there.
We therefore conclude that the one-magnon mode fully decays into the continuum at this momentum by quantum fluctuations.
This behavior is consistent with recent neutron scattering experiments and matrix product state calculations reported in Ref.~\cite{Hernandez2025}.
While the SAC analysis provides clear evidence for magnon decay, it does not by itself reveal the microscopic mechanism underlying this phenomenon. A deeper understanding of the decay requires complementary approaches, which we address in the following by discussing our results using SE and \gls{cst}.

\begin{figure}
    \centering
    \includegraphics[width=1\linewidth]{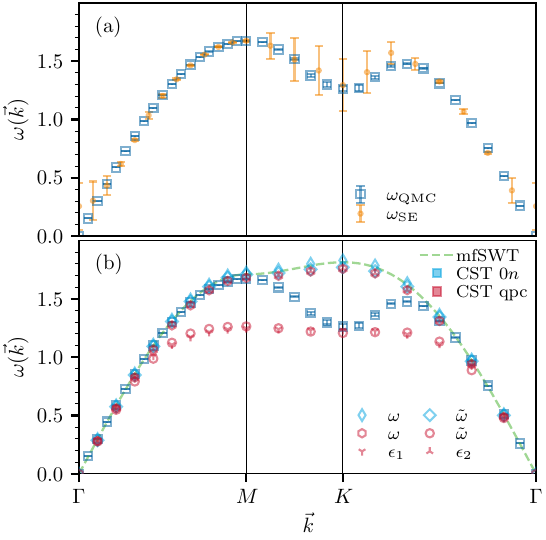}
    \caption{
    Panel (a) shows the comparison between the one-magnon dispersion $\omega_\mathrm{SE}$ from SE and the lower band edge of the dynamic structure factor $\omega_\mathrm{QMC}$ from QMC. The error bars of $\omega_\mathrm{SE}$ reflect the uncertainty of the averaged Pad\'{e} extrapolants.
    Panel (b) shows the dispersions from different evaluations in the \gls{cst} framework 
    for a linear system size of ${L=18}$ with periodic boundary conditions.
    The initial mean-field dispersion before the flow (mfSWT) is shown as a dashed green line.
    The \gls{cst} data stems from the $0n$ flow (blue markers) and the \gls{qpc} flow 
    stopped at $\ell=1$ (red markers), both with ($\tilde{\omega}$) and without ($\omega$) a re-diagonalization in the one-three magnon space.
    The red tripod markers mark the two lowest eigenvalues $\epsilon_{1,2}$ of the two-magnon 
    subspace with $S^z=0$ calculated from the stopped \gls{cst} flow.
    For comparison, the QMC dispersion $\omega_{\mathrm{QMC}}$ is also displayed.
    Note that due to the convergence issues, the \gls{cst} data is less reliable than the QMC and SE results.
    }
    \label{fig:qmc_se_cst_disp}
\end{figure}

In \cref{fig:qmc_se_cst_disp}a we compare the one-magnon dispersion $\omega_{\rm SE}$ from SE to the lower band edge of the dynamic structure factor $\omega_\mathrm{QMC}$ from QMC. The displayed SE data correspond to the average of Pad\'{e} approximants as described in \cref{sssec:pade} and the plotted error bar to the standard deviation reflecting the uncertainty. 
Overall, we find quantitative agreement throughout the Brillioun zone confirming that the lower band edge detected by QMC corresponds to the one-magnon excitation energies with a finite spectral weight in the dynamic structure factor.
However, there are two regions where the Pad\'{e} extrapolation has a large uncertainty, which can be understood as follows. 
First, at the $\Gamma$-point we expect a gap-closing transition in the XXZ model with a power-law behaviour $(1-\lambda)^{1/2}$ close to the Heisenberg point. 
This power-law behavior cannot be described by Pad\'{e} extrapolation and therefore the uncertainty becomes large close to the $\Gamma$-point.  
Second, close to the $K$-point we observe a large uncertainty, which is also expected if the magnons decay and the corresponding quasi-particle picture breaks down.
Although uncertainties in the Pad\'{e} extrapolation can also have different, unphysical origins, the SE results are fully consistent with the findings of QMC.

Next, we present the \gls{cst} results to corroborate the QMC results and to
provide insight in the decay mechanism at the $K$-point. 
As already mentioned, we observe divergent flows with the \gls{qpc} generator, which is consistent 
with the observation of a vanishing single-particle weight 
since energetically overlapping magnon subspaces cannot be untangled. 
To extract the single-particle dispersion we resort to the different approaches discussed in \cref{sssec:convergence}. 
All results are obtained using periodic boundary conditions and a linear system size of $L=18$.

First, we analyze the effective Hamiltonian obtained with the $0n$ generator.
Both dispersions, be it the bare dispersion $\omega_{0n}$ or the one from 
the re-diagonalization of the one-three magnon subspace $\tilde{\omega}_{0n}$, 
show marginal renormalization in the sense that the obtained dispersions display
no fundamental discernible differences compared to the mean-field spin-wave theory (mfSWT) 
dispersion of the initial Hamiltonian before the flow, see \cref{fig:qmc_se_cst_disp}b. 
Consequently, we do not observe any change in the position of the local minimum from $M$ to $K$ 
as found by QMC and SE, also displayed in \cref{fig:qmc_se_cst_disp}b.

With the flow of the $0n$ generator showing no significant effect, we analyze the \gls{qpc} generator stopped at the flow parameter $\ell=1/J$.
We find again that the bare dispersion $\omega_{\mathrm{qpc}}$ after the stopped \gls{qpc} flow shows 
only marginal differences compared to the initial mfSWT dispersion.  
However, the re-diagonalized dispersion $\tilde{\omega}_{\mathrm{qpc}}$ displays a 
significant renormalization in the high-energy sector compared to all previous results obtained by \gls{cst}.
Namely, in the high-energy region, the dispersion $\tilde{\omega}_{\mathrm{qpc}}$ is broadly lowered to a value close to the minimum of the QMC results at the $K$-point.
However, compared to the QMC results, the dispersion rather forms a plateau 
between the $M$- and the $K$-point.
Therefore, both dispersions  differ significantly at the $M$-point although they agree at the $K$-point.

How can a lower continuum edge of three-magnons arise? The only
possibility is that binding occurs. One option is that a genuine three-magnon bound state arises.
Yet this is highly non-trivial  though it may appear, at least in low-dimensional systems \cite{schmi22b}.
The more natural hypothesis is that two-magnon bound states arise since a strong
attractive magnon-magnon interaction is established in the square lattice Heisenberg model \cite{Powalski2015,Powalski2018}.
In order to pursue this idea, we also considered the two-magnon subspace.
In particular, we analyze the subspace with $S^z=0$ because in this subspace 
 magnon-magnon interactions are the strongest.
 
\cref{fig:qmc_se_cst_disp}b shows the two lowest eigenvalues $\epsilon_{1,2}$ of the
two-magnon subspace.
We find clear evidence that the lowest eigenvalue $\epsilon_1$ is connected to a two-magnon 
bound state as it lies well below the bare dispersion $\omega_\mathrm{qpc}$ whereas $\epsilon_2$ coincides with $\omega_\mathrm{qpc}$. 
Interestingly, the dispersion of the bound state $\epsilon_1$ coincides with the dispersion 
$\tilde{\omega}_{\mathrm{qpc}}$, which gives a strong evidence that 
$\tilde{\omega}_{\mathrm{qpc}}$ in the area around $M$ and $K$ consists of excitations of three magnons instead of a single magnon.
Thus, the true single-magnon dispersion merges with the continuum as it is not the 
lowest lying energy in this region anymore, which stands in contrast to the observation made for the Heisenberg model on a square lattice.
This observation perfectly aligns with the divergence of the \gls{qpc} generator, as it is not possible to decouple the different magnon subspaces, resulting in a divergent flow.  

When considering a stopped flow, we observe that the attempt to rotate the non-particle-conserving terms, especially scattering terms, to zero leads to strong interaction effects between magnons.
If the flow is not stopped, these effects will ultimately result in the divergence of the \gls{qpc} flow.
Although the stopped \gls{cst} captures the height and the formation of the local minimum at the 
$K$-point quantitatively, the formation of the plateau between $M$- and $K$-point 
appears to be inaccurate because QMC predicts a notably higher maximum.
We suspect that errors stemming from the truncation and the
unachievable endeavor of the \gls{qpc} generator to untangle the 
different magnon spaces lead to this unphysical plateau.
Thus, we consider the large extent of the bound state and the resulting plateau in the \gls{cst} dispersion as artifact as they are not supported by the other methods. 
The clear binding tendency provides evidence that the mechanism of magnon decay is related to bound states. 
We stress that the mere fact that the particle-conserving generator yields a diverging flow indicates that three-magnon states are shifted below one-magnon states energetically supporting this scenario of decay.
A generator scheme which distinguishes the $1:3$ processes according to the momentum
suggests itself to explore the plateau further by an improved  flow. But this
exceeds the scope of this article.

We also analyzed the behavior of this plateau in dependence on the flow parameter $\ell$.
The plateau forms early in the flow and its height decreases overall as $\ell$ increases. 
It is particularly impressive that the height aligns well with that of QMC results 
at the $K$-point when the previously defined stop condition for the flow is met. 

All in all, we find that the observed vanishing of the single-magnon weight in the QMC results at the $K$-point perfectly fits the observation made within the \gls{cst} framework.
Thus, a likely bound state in the two-magnon sector may lead to this vanishing weight.
It is precisely the correct treatment of the scattering terms that is responsible for the 
formation of the bound state since the bound state is not found using the $0n$ generator. 
However, interaction terms might be overestimated within the \gls{cst}, 
leading to a bound state in the whole higher-energy regime. 
In contrast, QMC and SE only see its effect close to the $K$-point.

\section{Conclusion}
\label{sec:conclusion}

In this work we investigate the physical properties of elementary magnon excitations above the long-range ordered ground state of the antiferromagnetic Heisenberg model on the honeycomb lattice. 
We focus on the one-magnon dispersion and the associated spectral weight of the dynamic structure factor at high energies near the corner of the Brillouin zone (the $K$-point).
Here dramatic deviations from linear spin-wave theory occur due to strong quantum fluctuations, which are even enhanced compared to the closely related Heisenberg model on the square lattice due to the smaller coordination number, and we find strong numerical evidence for a complete quantum decay of magnons confirming the scenario suggested in Ref.\,\cite{Hernandez2025} and amending it by the interpretation that it originates from magnon-magnon binding.

We employ three complementary methods: quantum Monte Carlo simulations, series expansion, and continuous similarity transformations. 
From QMC simulations we extract imaginary-time correlation functions and obtain spectral functions (the dynamic structure factor) by stochastic analytic continuation (SAC). 
Using a restricted-sampling scheme, we separate the weight of the magnon peak, 
which corresponds to the first macroscopic delta peak, from the remaining continuum. 
By optimizing the weight of this peak through the SAC cost function 
we determine the optimal one-magnon weight.
We find that the one-magnon weight becomes very small in a region around the $K$-point, while it remains finite at other momenta such as the $M$-point. 
At the $K$-point almost all spectral weight is transferred to the continuum above the magnon. 

A finite-size analysis based on several system sizes shows that the one-magnon weight at the $K$-point converges to zero, and that the lowest excitation energy is strongly reduced. 
We interpret this behavior as a full quantum decay of the magnon into the continuum at this momentum.
The SE results for the one-magnon dispersion, obtained by extrapolating the series from the Ising limit 
to the Heisenberg point in the XXZ model, support this conclusion. 
They show quantitative agreement with the excitation energies extracted from QMC simulations 
in large parts of the Brillouin zone. However, the extrapolation shows significant
uncertainties near the $K$-point, which again suggests a quantum decay of magnons
in this  region of momentum.

The \gls{cst} analysis in momentum space provides further confirmation of this scenario 
and yields additional insight into the mechanism behind the quantum decay. 
When we enforce a magnon quasi-particle description within \gls{cst}, the flow diverges.
This behavior is not observed in the same model on the square lattice.
The divergence indicates a breakdown of the magnon quasi-particle picture.
Plausibly, this results from the decay of single magnons into continua of multi-magnon states.
For the latter to become relevant, they must be shifted downwards in energy
because without such a shift of spectral weights the influence of multi-magnon continua
appears to be unimportant \cite{Gu2022}.
To analyze this more in depth, we stopped the flow before the divergence occurs.
Although the resulting dispersion shows discrepancies to the other methods, the \gls{cst} provides us with evidence for the underlying physical mechanism.
Within the effective model we indeed identify a two-magnon bound state below the single magnon dispersion.
The bound state appears in a large range of the Brillouin zone except near the $\Gamma$-point, but this is not supported by QMC for most areas of the Brillouin zone.
Only around the $K$-point its energy coincides with the excitation energies extracted from QMC simulations.
Thus, we attribute the quantum decay to a binding phenomenon of magnons. 

Overall, the three methods yield a consistent picture of the quantum decay of the one-magnon mode in the Heisenberg antiferromagnet on the honeycomb lattice lose to the $K$-point.
The results are in good agreement with previous theoretical and experimental studies of this system \cite{Maksimov2016, Ferrari2020, Wessler2020, Sala2021, Sala2023, Hernandez2025}.
It is important to emphasize that the decay arises solely due to quantum fluctuations 
because the model contains neither geometric frustration nor competing interactions.
The effect of quantum fluctuations is enhanced in the honeycomb lattice 
compared to the square lattice because of its lower coordination number.
It remains an intriguing open question whether a similar decay occurs in other lattices with low coordination numbers or whether it is restricted to the honeycomb geometry.

\section{Acknowledgements}
We thank Max H\"ormann for developing and formulating the initial description of the PCAT-based linked-cluster methodology, for performing the corresponding calculations, and for providing the high-order series used in this work.
We thank Andreas Läuchli for fruitful discussions and Jan Alexander Koziol and Anja Langheld for their comments to the manuscript.
We gratefully acknowledge financial support by the Deutsche Forschungsgemeinschaft (DFG, German Research Foundation) through projects UH 90-14/2 (GSU), and SCHM 2511/13-2 (KPS). We thankfully acknowledge HPC resources provided by the Erlangen National High Performance Computing Center (NHR@FAU) of the Friedrich-Alexander-Universit\"at Erlangen-N\"urnberg (FAU). KPS acknowledges further financial support by the Munich Quantum Valley, which is supported by the Bavarian state government with funds from the Hightech Agenda Bayern Plus. 

\section{Data availability}
The data that support the findings of this article are openly available \cite{RawData}, embargo periods may apply.

%

\end{document}